\def\year{2019}\relax
\documentclass[letterpaper]{article} 
\usepackage{aaai19}  
\usepackage{times}  
\usepackage{helvet}  
\usepackage{courier}  
\usepackage{url}  
\usepackage{graphicx}  

\usepackage{booktabs}
\usepackage{multirow}
\usepackage{color}
\usepackage{amsmath}
\usepackage{amsfonts}
\usepackage{subfigure}
\usepackage{bm}

\begin{document}
%
\title{TS-CNN: Text Steganalysis from Semantic Space Based on Convolutional Neural Network}
\author{Zhongliang Yang, Nan Wei, Junyi Sheng, Yongfeng Huang, Yu-Jin Zhang\\
Department of Electronic Engineering, Tsinghua University, Beijing, 100084, China.\\
National Key Laboratory for Novel Software Technology, Nanjing University, Nanjing 210023, China\\
College of Software Engineering, Sichuan University, Sichuan, 610065, China.\\
E-mail: yangzl15@mails.tsinghua.edu.cn, yfhuang@tsinghua.edu.cn\\
}
\maketitle
\begin{abstract}
 
Steganalysis has been an important research topic in cybersecurity that helps to identify covert attacks in public network. With the rapid development of natural language processing technology in the past two years, coverless steganography has been greatly developed. Previous text steganalysis methods have shown unsatisfactory results on this new steganography technique and remain an unsolved challenge. Different from all previous text steganalysis methods, in this paper, we propose a text steganalysis method(TS-CNN) based on semantic analysis, which uses convolutional neural network(CNN) to extract high-level semantic features of texts, and finds the subtle distribution differences in the semantic space before and after embedding the secret information. To train and test the proposed model, we collected and released a large text steganalysis(CT-Steg) dataset, which contains a total number of 216,000 texts with various lengths and various embedding rates. Experimental results show that the proposed model can achieve nearly 100\% precision and recall, outperforms all the previous methods. Furthermore, the proposed model can even estimate the capacity of the hidden information inside. These results strongly support that using the subtle changes in the semantic space before and after embedding the secret information to conduct text steganalysis is feasible and effective.

\end{abstract}

\section{introduction}

In the monograph on information security\cite{shannon1949communication}, Shannon summarized three basic information security systems: encryption system, privacy system, and concealment system. while encryption system and privacy system ensure information security, they also expose the existence and importance of information, making it more vulnerable to get attacks, such as interception and cracking\cite{Bernaille2007Early}. Concealment system is very much different from these two secrecy systems. It uses various carriers to embed confidential information and then transmit through public channels, hide the existence of confidential information to achieve the purpose of not being easily suspected and attacked\cite{Simmons1984The}. The biggest characteristic of steganography is the strong concealment of information. From another aspect, strong concealment of steganography can be used by hackers, terrorists, and other law breakers for malicious intent. Hence, to design an automatic steganography detection method becomes a very promising and challenging task.

\begin{figure}[!tp]
  \centering
  \includegraphics[width=\linewidth]{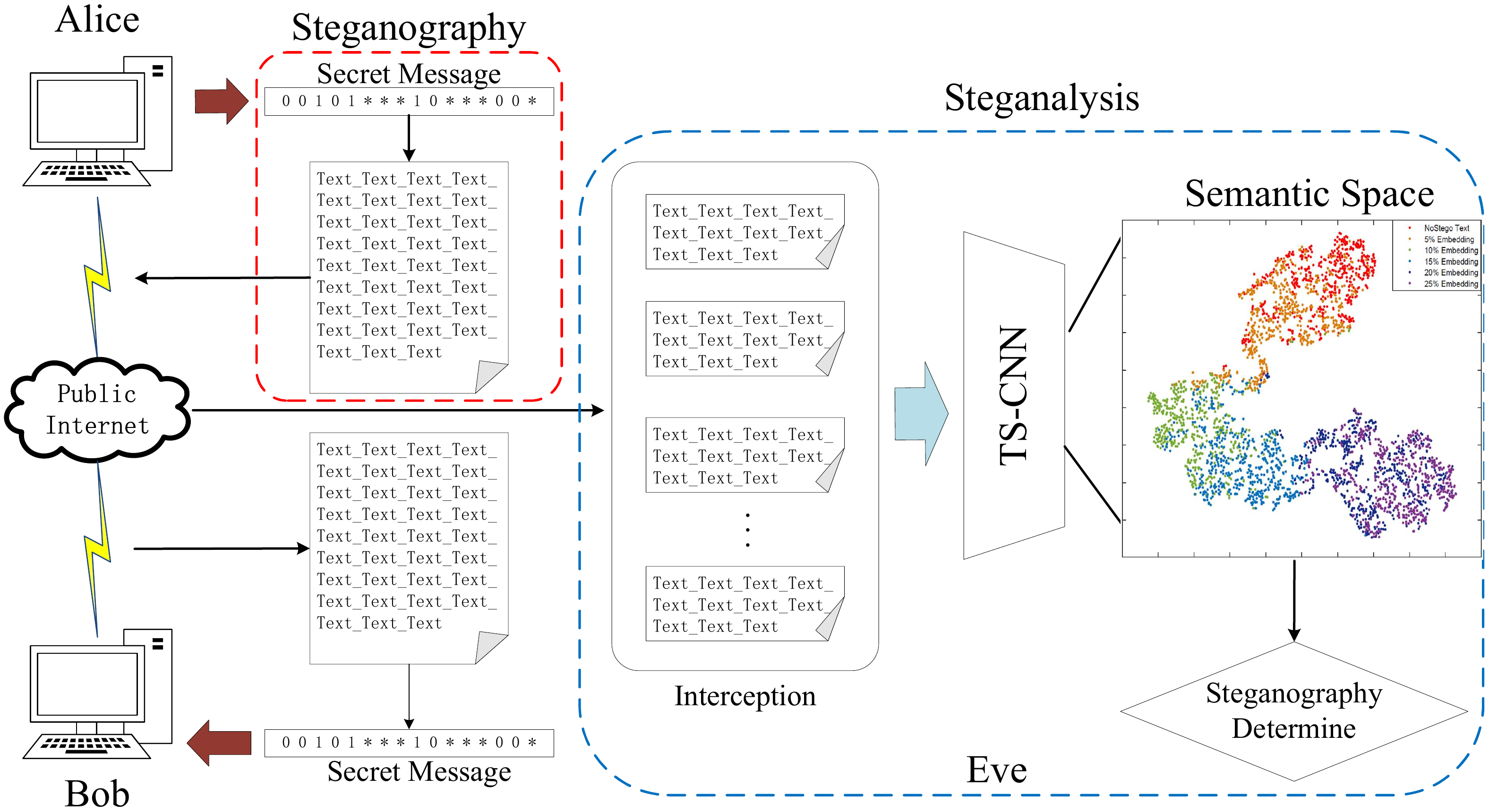}
  \caption{The overall framework of the text steganalysis technology(TS-CNN) proposed in this paper, which uses CNN to extract high-level semantic features of texts, and finds the subtle differences in the semantic space before and after embedding the secret information to achieve high efficiency and high accuracy detection of steganographic texts in the network.}
  \label{fig:1}
\end{figure}

Steganalysis is the technique that againsts steganography. From the perspective of steganography, the sender needs to ensure that the statistical distributions of the carrier do not change before and after hiding information. However, for steganalysis, the detector needs to find out as much as possible about the difference of statistical distribution between the covertext and the stegotext.

There are various media forms of carrier that can be used for information hiding, including image \cite{fridrich2009steganography}, audio \cite{yang2017sudoku}, text \cite{Luo2017Text,fang2017generating} and so on \cite{Cox2001The}. Among them, text is the most widely used information carrier in daily life. In recent years, more and more text based information hiding methods have emerged\cite{Desoky2010Comprehensive,Luo2017Text,fang2017generating}. Previous works on text steganography can be divided into two big families: format based method \cite{Zou2005Formatted} and content based method \cite{Bennett2004Linguistic}. Text format based information steganography usually treats text as a specially coded image, most of them achieve information hiding by modifying the appearance of the document, like inter-character space \cite{Chotikakamthorn1998Electronic}, word-shifting \cite{Shirali2006A}, character-coding \cite{Low1998Document}, etc. Text content based methods mainly use Natural Language Processing technology \cite{Desoky2010Comprehensive} to modify text content to hide information. They implement information hiding by replacing some of the words in the sentences \cite{Mahato2017A}, or by changing their syntactical structure \cite{Murphy2007The}. Such methods need to ensure that the modified text satisfies the requirements of semantic correctness and grammatical rationality. 

Most of the exist text steganalysis methods construct statistical analysis features based on text structures, lexical collocations, language models, etc., hoping to find the difference of their statistical distribution between the covertext and the stegotext \cite{Meng2009Linguistic,samanta2016real,Yu2012Steganalysis,din2015performance}. These methods only analysis a certain aspect of the carrier, thus have great limitations.

In recent years, with the development of deep neural network technology in Natural Language Processing, more and more steganographic methods based on automatic text generation have emerged, which is also called coverless steganography \cite{Liu2016A,fang2017generating,Luo2017Text}. Through some natural language processing methods, they automatically generate a piece of text, and finally achieve information hiding by properly encoding the words during text generation. The advantage of coverless steganography is that it does not need to be given a carrier in advance, but can automatically generate a piece of text according to the covert information. In the process of generation, it can learn the statistical model of a large number of samples and generate a steganographic carrier that conforms to its statistical distribution, so it can achieve high concealment and high-capacity information hiding at the same time. Traditional text steganalysis methods gradually fails in the face of this new steganography method, which brings great security risks to cyberspace security. 

In this paper, we propose a text steganalysis method (TS-CNN) based on semantic analysis, which uses convolutional neural network(CNN) to extract high-level semantic features of texts, and finds the subtle distribution differences in the semantic space before and after embedding the secret information. Compared with the previous text steganographic analysis method, the biggest feature of our method is that we do not need to construct the corresponding steganographic analysis features, but use the powerful feature extraction capabilities of convolutional neural networks to automatically extract high-level semantic features of text through extensive training. Then we map it to the high-dimensional semantic space and make a steganographic decision by analyzing the nuances of the semantic space distribution before and after the steganography. The framework of the proposed model has been shown in Figure \ref{fig:1}.

\section{Related work}

Previous text steganalysis algorithms can be divided into two categories: document appearance analysis based and natural language statistical analysis based. Steganalysis methods based on document appearance analysis are mainly aimed at format based steganography. For example, L.Li \emph{et al.}\cite{li2008statistical} presented a statistical analysis of a kind of word-shift text-steganography by using neighbor difference (length difference of two consecutive spaces). In addition, a large number of text steganalysis methods are based on natural language statistical properties. For example, Yang \emph{et al.}\cite{Yang2010Linguistic} proposed a novel linguistics steganalysis approach based on meta features and immune clone mechanism. They defined 57 meta features to represent texts, including the average length of words, the space rate, the percentage of letters and so on. Then the immune clone mechanism was exploited to select appropriate features so as to constitute effective detectors. Meng \emph{et al.}\cite{Meng2009Linguistic} proposed a linguistic steganography detecting algorithm using Statistical Language Model(SLM). They computed the perplexity of normal text and stego-text with the language model, and then they determined whether the text inputted contains covert information by setting a threshold. Taskiran \emph{et al.}\cite{Taskiran2006Attacks} trained a support vector machine (SVM) classifier based on the statistical features obtained from the language models. Finally, they classified a given text has secret information inside or not based on the output of the SVM classifier. Samanta \emph{et al.} \cite{samanta2016real} proposed statistical text steganalysis tools based on Bayesian Estimation and Correlation Coefficient methodologies. Din \emph{et al.}\cite{din2015performance} proposed a formalization of genetic algorithm method in order to detect hidden message on an analyzed text. In paper\cite{chen2008linguistic}, a novel statistical algorithm for linguistic steganography detection was presented. They used the statistical characteristics of correlations between the general service words gathered in a dictionary to classify the given text segments into stego-text segments and normal text segments. 

These above steganalysis methods can achieve good detection effect for certain specific steganographic algorithms, but they show unsatisfactory results for the coverless text steganography. The steganalysis method proposed in this paper is very different from the methods described above. Our thoughts is that when embed secret information in texts, the embedded information will definitely affect the expression of text semantics and its distribution in the semantic space. By detecting the difference in the semantic space distribution, we can finally achieve steganographic detection for coverless text steganography.

\section{Coverless Text steganography(CT-Steg) Dataset}

\begin{table*}[ht]
  \renewcommand\arraystretch{1.5}
  \centering
  \begin{tabular}{c|c|c|c|c|c|c|c|c}
  \toprule[1.5pt]
  \multicolumn{2}{c|}{Embedding Rate} &0\% &5\% &10\% &15\% &20\% &25\% & Total\\
  \hline
  \multirow{2}{*}{Five-Words(FW)} &{Four Lines(FL)} &4,000 &10,000 &10,000 &10,000 &10,000 &10,000 &54,000\\
  \cline{2-9}
  &{Eight Lines(EL)} &4,000 &10,000 &10,000 &10,000 &10,000 &10,000 &54,000\\
  \hline
  \multirow{2}{*}{Seven-Words(SW)} &{Four Lines(FL)} &4,000 &10,000 &10,000 &10,000 &10,000 &10,000 &54,000\\
  \cline{2-9}
  &{Eight Lines(EL)} &4,000 &10,000 &10,000 &10,000 &10,000 &10,000 &54,000\\
  \hline
  \multicolumn{2}{c|}{Total} &16,000 &40,000 &40,000 &40,000 &40,000 &40,000 &216,000\\
  \bottomrule[1.5pt] 
  \end{tabular}
  \caption{\label{tab:1}The collected large text steganalysis(CT-Steg) dataset, which are generated by the above model proposed in \cite{Luo2017Text}. It has a total number of 216,000 poems including confidential information with various embedded rates under various formats.}
\end{table*}

To the best of our knowledge, there is no public steganalysis dataset available. In order to test our method and also to promote the development of related fields, in this study, we collected and constructed a large-scale coverless text steganography(CT-Steg) dataset, which contains various lengths of steganographic texts at various embedding rates. 

Text samples with covert information in CT-Steg are generated by currently state of the art text steganography algorithm proposed in \cite{Luo2017Text}. Considering the randomness of natural text generation, Luo \emph{et al.} \cite{Luo2017Text} uses special formatted text (poem) as a vehicle for automatic generation and steganography. The benefit of such texts have specific formats and templates so that the generation process is more controllable. Therefore, the generated steganographic text has stronger concealment and becomes the most powerful text steganography algorithm at present.

\begin{figure}[ht]
\centering
\includegraphics[width=0.8\linewidth]{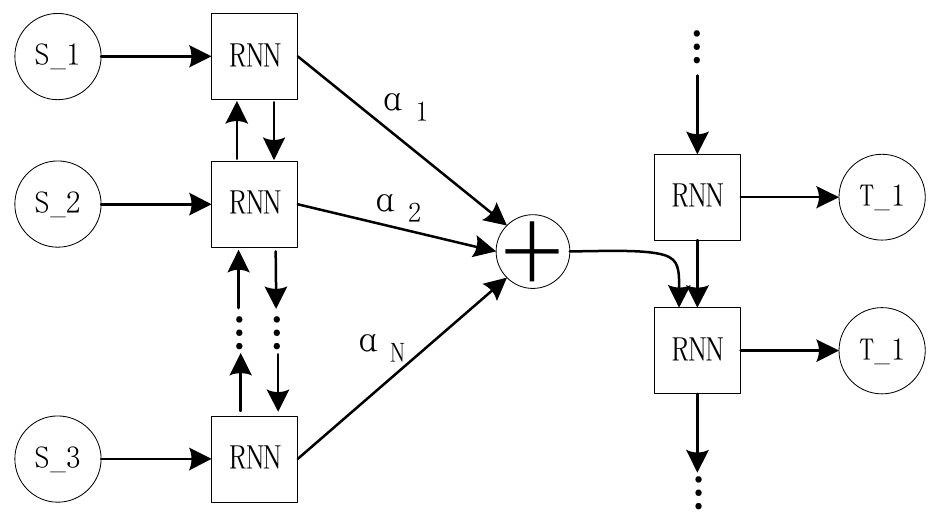}
\caption{Generation method proposed in \cite{Luo2017Text}. They used bidirectional recurrent neural network(RNN) with attention mechanism to ensure the correlation between sentences.}
\label{fig:2}
\end{figure}

Figure \ref{fig:2} shows the generation framework proposed in \cite{Luo2017Text}. Since classical Chinese quatrains have strong semantic relevance between adjacent lines, they use an attention-based bidirectional encoder-decoder model to build the line-to-line module. The encoder first maps the input sentence to a feature vector which contains the semantic meaning of each character of the input. The decoder calculates and generates the next sentence based on the feature vector of this input sentence. The attention mechanism calculates the importance of each word in the input sentence when understanding the previous sentence and then generating the next one. In this way, the semantic continuity of each sentence in the generated poem can be guaranteed.

The CT-Steg dataset has a total number of 216,000 poems including confidential information with various embedding rates under various formats, which have been shown in Table \ref{tab:1}. In CT-Steg dataset, poetry can be divided into two major categories, one is five-word per line, and the other is seven-word per line. They can be further divided into poetry that contains four lines or eight lines. The embedding rate ranges from 0\% to 25\%, 0\% representing no covert information in the poems we generated. Since the size of the candidate pool in each part of the poem is 0 when the embedding rate is 0\%, the resulting poem is fixed when the first word is given. Therefore, the number of poems that can be generated is limited, and the original paper\cite{Luo2017Text} only provides 4000 initial states, so only 4000 poems can be generated. Under other embedding rates, we have generated 10,000 poems, so in the end the CT-Steg dataset contains 216,000 poems as shown in Table \ref{tab:1}. Our experiments are based on this dataset, including the training and testing of the model. But while actually, our algorithm can also be directly applied to other text steganographic method.

\section{TS-CNN Methodology}

Benefiting from the tremendous development in deep neural network field in the past two years, there has been relevant researches on the use of CNN to extract high-level semantic features of texts\cite{Mikolov2013Linguistic,kim2014convolutional}. For each input sentence $S$, we can illustrate it with a matrix $\mathrm{D} \in \mathbb{R}^{N \times K}$, where the $i$-th row indicates the $i$-th word in passage $S$, each word is represented as a $K$-dimension vertor. Generally, let $\mathrm{X}_{i:j}$ refer to the matrix which consists of the words vectors of the $i$-th word to the $j$-th word. The width of each convolution kernel is the same as the width of the input matrix. If the height of the convolution kernel is $h$, and the convolutional kernel is expressed as $\mathrm{W} \in \mathbb{R}^{h \times K}$, then the feature $c_i$ extracted from $\mathrm{X}_{i:j}$ by the convolutional kernel can be:

\begin{equation}
{c_i} = f(\mathrm{W} \cdot \mathrm{X}_{i:j}+b_i),
\end{equation}

\noindent where $f$ is a nonlinear function and $b_i$ is the bias. The convolutional kernel slides from the top to the bottom of the text matrix, which produces a feature map named as $\mathbf{c}$:

\begin{equation}
\mathbf{c} = [c_1,c_2,...,c_{N-h+1}].
\end{equation}

We then apply a pooling layer, and the pooling can be maximum pooling and average pooling. If it is the maximum pooling we can take the maximum value as $\widetilde{c} = \max\{\mathbf{c}\}$. The process described above is a process in which one convolution kernel produces one feature. The actual model has multiple convolution layers in the calculation, and each layer has many convolution cores. The generation of each feature is described as above. Let the feature vector produced at the last layer be:

\begin{equation}
\mathbf{z} = [\widetilde{c}_1,\widetilde{c}_2,...,\widetilde{c}_m].
\end{equation}

\begin{figure}[ht]
  \centering
  \subfigure[TS-CNN(Single)]{
      \label{fig:subfig:a} 
      \includegraphics[width=\linewidth]{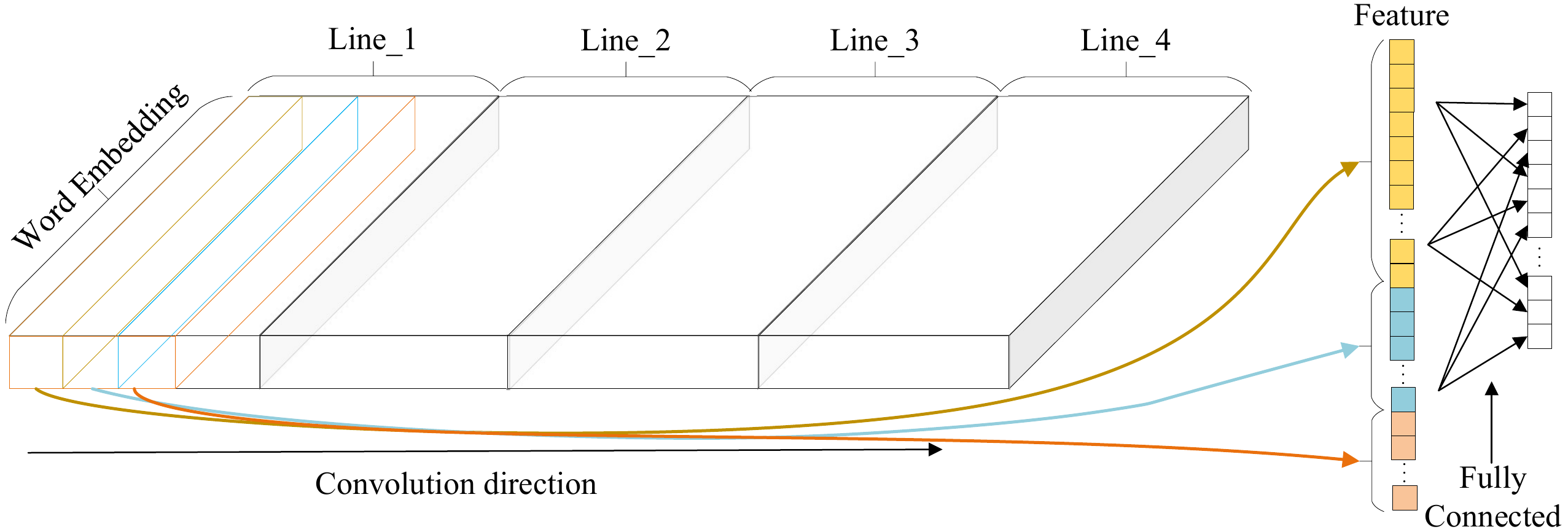}}
    \hspace{1in}
  \subfigure[TS-CNN(Multi)]{
      \label{fig:subfig:b} 
      \includegraphics[width=\linewidth]{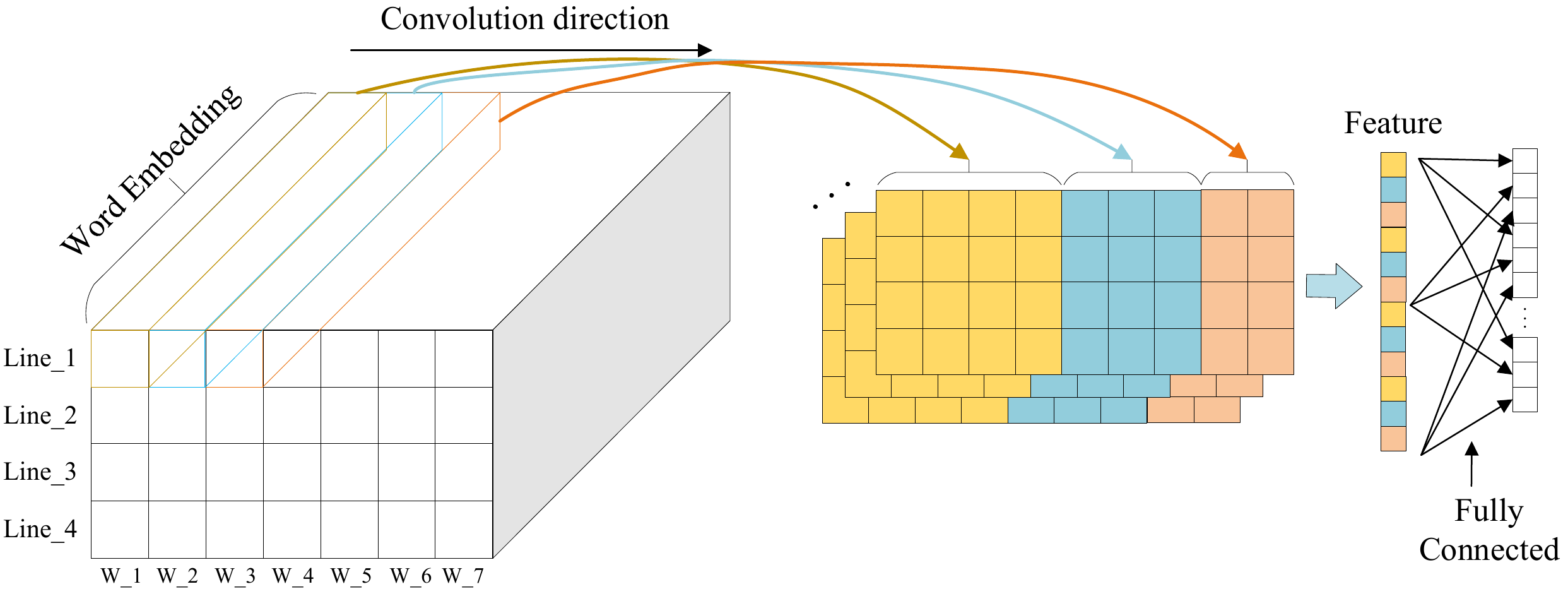}}
    \caption{The proposed two corresponding models for feature fusion and classification, called TS-CNN(Single) and TS-CNN(Multi).}
  \label{fig:4}
\end{figure}

Each poetry contains more than one sentences. In general, each line can be seen as a single sentence, and there is a structural and semantic correspondence between each sentence. For this situation, we propose two corresponding models for feature fusion and classification, called TS-CNN(Single) and TS-CNN(Multi), and their structures have been shown in Figure \ref{fig:4}. 

The TS-CNN(Single) model concatenates each line of each poem inputted, transforms the entire poem into one single sentence, and then uses the model described above to perform feature extraction. The extracted feature vector is the feature expression of the entire poem. The TS-CNN(Multi) model treats each line of each poem as a separate sentence and then the extracted characteristics are adjusted and rearranged to form the feature vector of the entire poem. This design makes TS-CNN(Single) pays more attention to the semantic coherence between sentences, while TS-CNN(Multi) pay more attention to the expression of the internal semantics of each sentence. Without causing ambiguity, here we use $\mathbf{z}$ to represent the feature vector extracted for both of these two models. 

By defining a weight matrix $\mathrm{W}_F$, we compute the weighted sum of each feature element:

\begin{equation}
y = \mathrm{W}_F \mathbf{z} + b_f,
\end{equation}

\noindent where $\mathrm{W}_F$ and $b_f$ are learned weight matrix and bias, the values in weight matrix $\mathrm{W}_F$ reflect the importance of each feature. To get normalized output between $[0, 1]$, we put the value through a sigmoid function $S$, and the final output is 

\begin{equation}
O = S(y) = S(\mathrm{W}_F\mathbf{z} + b_f).
\end{equation}

The output value reflects the probability that our model believes that the input text contains confidential messages. We can set a detection threshold, then the final detection result can be expressed as

\begin{equation}
Result = 
\left\{
\begin{aligned}
& Stegotext & (O \geq threshold) \\
& Covertext & (O < threshold)
\end{aligned}
\right.
\end{equation}

The loss function of the whole network consists of two parts, one is the error term and the other is the regularization term, which can be described as:

\begin{equation}
  \begin{aligned}
  loss &= \sum_{num}(\mathrm{P}-\mathrm{T})^\top(\mathrm{P}-\mathrm{T}) + \|\mathrm{W}_F\|_2 \\
  &= \sum_{num}\sum^L_{i=0}(\mathrm{P}_i-\mathrm{T}_i)^2 + \|\mathrm{W}_F\|_2,
  \end{aligned}
\end{equation}

\noindent where $num$ is the batch size of texts, $\mathrm{P}$ indicates the output of the classifier, and each element as $\mathrm{P}_i$ represents the probability that the $i$-th sample is judged to contain covert information. $\mathrm{T}$ is the actual label of input sample. The error term in the loss function calculates the mean square error (MSE) between the prediction vector and the actual label. In order to strengthen the regularization and prevent overfitting, we adopted the dropout mechanism \cite{krizhevsky2012imagenet} and a constraint on $L_2$ norms of the weight vectors during the training process.

\section{Experiments and Analysis}

In this section, we designed several experiments to test the proposed model. We will first introduce the model structure and experimental parameter settings.

\subsection{Experimental Setting and Training Details}

The structure of the convolutional neural network used in this study is as follows: an embedding layer with randomly initialized word vectors of 400 dimensions; a convolutional layer with three different sizes of convolutional kernels with height 1, 2, 3, and each size has 128 convolution kernels; an average pooling layer and a fully connected layer following with a softmax classifier. We set the detection threshold to be 0.5. All the values of these super-parameters are decided through comparison experiments. We use SGD with momentum 0.9 to train the parameters of our network.

For each training and testing on TS-CNN, we randomly pick up 3,500 positive and 3,500 negative samples from Nostego-texts(embedding rate: 0\%) and Stego-texts with different embedding rate, and then randomly pick up another 500 from each dataset for model testing.

\subsection{Evaluation Results and Discussion}

\subsubsection{Steganalysis Efficiency}

We first test the efficiency of the proposed model, which is the time it takes to analyze a piece of text. Results in Table \ref{tab:2} further validate the efficiency of our model for steganalysis. For the shortest text in the dataset FW-FL, where each of them contains 20 words, only takes an average time of 4.81/4.76ms for TS-CNN(Single) and TS-CNN(Multi). As the sample length increases, the required steganalysis time gradually increases. Even so, for the longest text in SW-EL, which contains 56 words per sample, it takes 9.78/9.78ms on average.

\begin{table}[ht]
  \renewcommand\arraystretch{1.5}
  \centering
  \begin{tabular}{cc|c|c}
  \toprule[1.5pt]
  \multicolumn{2}{c|}{Model} &TS-CNN(Single) &TS-CNN(Multi)\\
  \hline
  \multirow{2}{*}{FW} &{FL} &4.81$\pm$1.16(ms) &4.76$\pm$0.56(ms)\\
  \cline{2-4}
  &{EL} &7.48$\pm$1.41(ms) &7.50$\pm$0.65(ms)\\
  \hline
  \multirow{2}{*}{SW} &{FL} &6.20$\pm$1.30(ms) &6.52$\pm$0.62(ms)\\
  \cline{2-4}
  &{EL} &9.78$\pm$1.46(ms) &9.78$\pm$0.98(ms)\\
  
  \bottomrule[1.5pt] 
  \end{tabular}
  \caption{\label{tab:2}The average prediction time for each sample in test set.}
\end{table}

\subsubsection{Steganalysis Accuracy}

In order to objectively and truly reflect the performance of TS-CNN, in this section, we conduct test on several representative text steganalysis algorithms. \cite{Meng2009Linguistic} use the trained language model to calculate the perplexity of each input sample text, and use them as the statistical features to determine whether they contain covert information. \cite{samanta2016real} proposed statistical text steganalysis tools based on Bayesian Estimation and Correlation Coefficient methodologies. \cite{din2015performance} proposed a formalization of genetic algorithm method in order to detect hidden message on an analyzed text. 

\begin{table*}[ht]
  \renewcommand\arraystretch{1.5}
  \centering
  \resizebox{\textwidth}{50mm}{
  \begin{tabular}{c|c|c|ccc|ccc|ccc|ccc|ccc}
  \toprule[2pt]
  \multicolumn{3}{c|}{Method} &\multicolumn{3}{|c|}{\cite{Meng2009Linguistic}} &\multicolumn{3}{|c|}{\cite{samanta2016real}}  &\multicolumn{3}{|c|}{\cite{din2015performance}} &\multicolumn{3}{|c|}{TS-CNN(Single)} &\multicolumn{3}{|c}{TS-CNN(Multi)}\\
  \hline
  \multicolumn{3}{c|}{Metric} &Acc &P &R &Acc &P &R &Acc &P &R &Acc &P &R &Acc &P &R\\
  \hline
  \multirow{10}{*}{FW} &\multirow{5}{*}{FL} &5\% &0.521 &0.597 &0.518 &0.768 &0.775 &0.768 &0.724 &0.706 &0.766 &0.907 &0.931 &0.872 &\textbf{0.929} &\textbf{1.000} &\textbf{0.877}\\
  &  &10\% &0.600 &0.671 &0.588 &0.868 &0.877 &0.868 &0.843 &0.840 &0.848 &0.955 &0.921 &\textbf{0.959} &\textbf{0.969} &\textbf{1.000} &0.877\\
  &  &15\% &0.681 &0.747 &0.659 &0.884 &0.896 &0.884 &0.895 &0.896 &0.894 &0.970 &0.942 &0.960 &\textbf{0.973} &\textbf{0.978} &\textbf{1.000}\\
  &  &20\% &0.705 &0.769 &0.681 &0.872 &0.885 &0.861 &0.899 &0.892 &0.908 &\textbf{0.990} &\textbf{1.000} &\textbf{1.000} &0.981 &0.979 &0.979\\
  &  &25\% &0.798 &0.858 &0.765 &0.868 &0.893 &0.864 &0.933 &0.930 &0.936 &\textbf{0.990} &0.962 &\textbf{1.000} &0.986 &\textbf{1.000} &0.958\\
  \cline{2-18}
  &\multirow{5}{*}{EL} &5\% &0.515 &0.598 &0.513 &0.775 &0.775 &0.775 &0.750 &0.778 &0.700 &0.815 &0.869 &\textbf{0.769} &\textbf{0.858} &\textbf{0.969} &0.718\\
  &  &10\% &0.592 &0.671 &0.579 &0.917 &0.956 &0.897 &0.904 &0.902 &0.906 &\textbf{0.986} &\textbf{1.000} &\textbf{0.980} &0.960 &0.956 &0.897\\
  &  &15\% &0.675 &0.742 &0.654 &0.918 &0.924 &0.918 &0.944 &0.932 &0.958 &\textbf{0.985} &0.952 &\textbf{0.975} &0.961 &\textbf{1.000} &0.901\\
  &  &20\% &0.712 &0.772 &0.689 &0.915 &0.948 &0.915 &0.961 &0.964 &0.958 &\textbf{0.999} &\textbf{1.000} &\textbf{1.000} &0.975 &0.963 &\textbf{1.000}\\
  &  &25\% &0.819 &0.869 &0.789 &0.911 &0.923 &0.901 &0.970 &0.974 &0.966 &\textbf{0.996} &\textbf{1.000} &\textbf{1.000} &0.987 &0.981 &0.981\\
  \toprule[1.2pt]
  \multirow{10}{*}{SW} &\multirow{5}{*}{FL} &5\% &0.539 &0.616 &0.533 &0.666 &0.667 &0.666 &0.710 &0.703 &0.726 &\textbf{0.922} &0.978 &\textbf{0.920} &0.917 &\textbf{1.000} &0.800\\
  &  &10\% &0.619 &0.691 &0.604 &0.898 &0.901 &0.898 &0.918 &0.927 &0.908 &0.980 &0.962 &0.927 &\textbf{0.981} &\textbf{0.979} &\textbf{0.942}\\
  &  &15\% &0.691 &0.762 &0.667 &0.928 &0.931 &0.941 &0.954 &0.960 &0.948 &0.984 &0.979 &\textbf{0.979} &\textbf{0.991} &\textbf{1.000} &0.941\\
  &  &20\% &0.730 &0.796 &0.703 &0.925 &0.930 &0.898 &0.973 &0.972 &0.974 &\textbf{0.997} &\textbf{1.000} &\textbf{1.000} &0.991 &\textbf{1.000} &0.979\\
  &  &25\% &0.810 &0.865 &0.779 &0.916 &0.927 &0.913 &0.985 &0.986 &0.984 &\textbf{0.995} &\textbf{1.000} &0.976 &\textbf{0.995} &\textbf{1.000} &\textbf{0.981}\\
  \cline{2-18}
  &\multirow{5}{*}{EL} &5\% &0.523 &0.624 &0.519 &0.656 &0.656 &0.656 &0.675 &0.659 &0.724 &0.861 &0.956 &0.745 &\textbf{0.893} &\textbf{0.974} &\textbf{0.796}\\
  &  &10\% &0.589 &0.682 &0.575 &0.926 &0.927 &0.926 &0.940 &0.956 &0.922 &\textbf{0.993} &\textbf{1.000} &\textbf{0.981} &0.972 &0.960 &0.941\\
  &  &15\% &0.651 &0.721 &0.633 &0.942 &0.942 &0.942 &0.950 &0.957 &0.942 &\textbf{0.989} &\textbf{1.000} &\textbf{1.000} &0.971 &\textbf{1.000} &0.962\\
  &  &20\% &0.696 &0.754 &0.675 &0.958 &0.959 &0.958 &0.978 &0.982 &0.974 &\textbf{0.995} &\textbf{1.000} &\textbf{1.000} &0.988 &0.977 &\textbf{1.000}\\
  &  &25\% &0.793 &0.829 &0.774 &0.954 &0.957 &0.952 &0.984 &0.984 &0.984 &\textbf{0.999} &\textbf{1.000} &\textbf{1.000} &0.995 &0.960 &\textbf{1.000}\\
  \bottomrule[2pt] 
  
  \end{tabular}}
  \caption{\label{tab:3}Results of different steganalysis methods.}
  
\end{table*}

We use several evaluation indicators commonly used in classification tasks to evaluate the performance of our model, which are precision, recall, F1-score and accuracy. Experiment results have been shown in Table \ref{tab:3}.

\begin{figure}[ht]
\centering
\includegraphics[width=\linewidth]{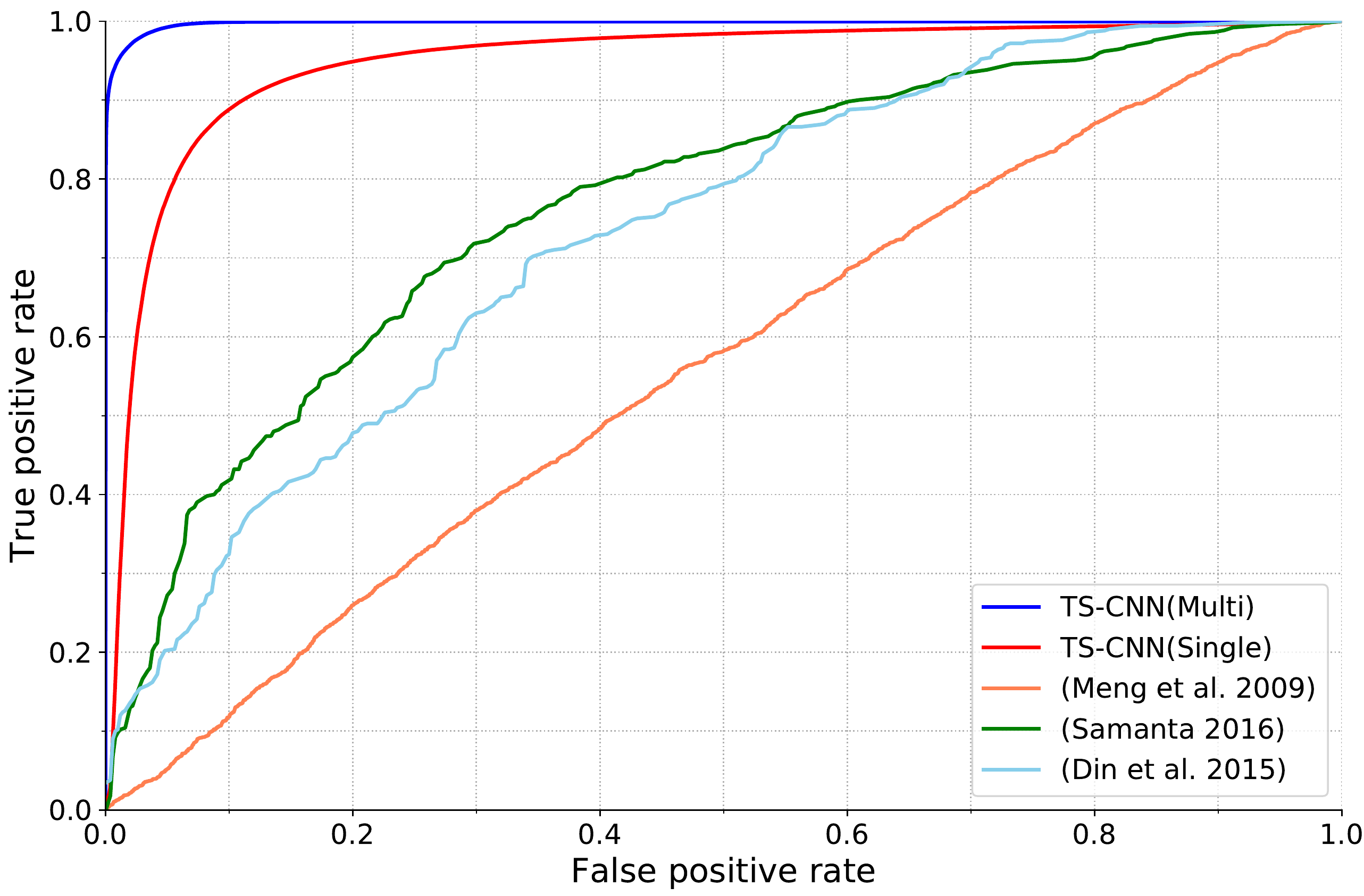}
\caption{The ROC curves of each model on SW-FL set at 5\% embedding rat. It's clear to see that the proposed models have obvious detection performance advantages over other models at low embeding rate.}
\label{fig:7}
\end{figure}

Compared to other text steganalysis methods, both TS-CNN (Single) and TS-CNN (Multi) have achieved the best detection results on various metrics, include different text lengths and different embedding rates. Especially in the case of low embedding rate, for example, when the embedding rate is only about 5\%, our model has more obvious detection performance advantages over others. We have also plotted the ROC curves of each model on SW-FL set at 5\% embedding rate, which is shown in Figure \ref{fig:7}. Figure \ref{fig:7} illustrates the clear advantage of the proposed model over other models. At the same time, the recall of our model is also high, which means that the probability of missed detection is extremely low in our model. Combined with these test results, we can find that our model has significantly practical value. 

\begin{figure}[ht]
\centering
\includegraphics[width=\linewidth]{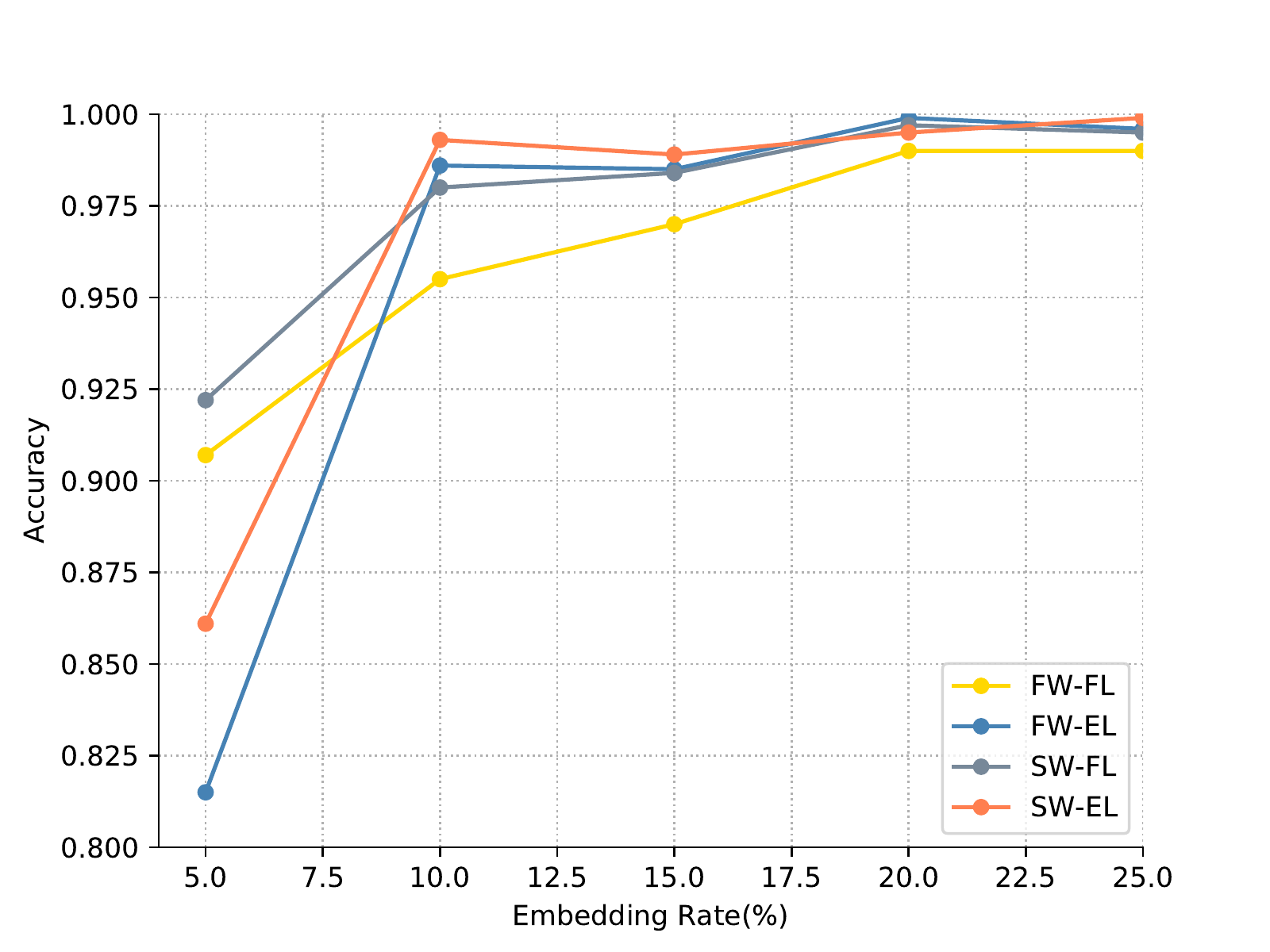}
\caption{The change of the steganalysis performance of the TS-CNN(Single) with the increase of embedding rate.}
\label{fig:8}
\end{figure}

Moreover, we noticed that the detection performance of each model has improved with the increase of the embedding rate. These results do meet our previous conjecture that with the increase of embedding rate, noise also increases, which will definitely damage the expression ability of texts semantics. Therefore it is easier to be distinguished from the unembedded information texts. Figure \ref{fig:8} shows the change of steganalysis performance of the TS-CNN(Single) with the increase of embedding rate. From Figure \ref{fig:8}, we notice that when the embedding rate is higher than 10\%, the detection accuracy rate stays at a very high level (above 97.5\%).

Text length is a significant performance impact factor, and the effect on the two proposed models is just the opposite. We can compare the results on EL and FL set under each embedding rate, since the length of each sample in EL is twice as long as that of FL. For the TS-CNN (Single), except for the condition of low embedding rate (ER=5\%), most of the test results on EL are better than those on FL. But for TS-CNN (Multi), most of the test results on FL are better than those on EL. Similarly, for the same pattern (FL or EL), the text length of the SW sample is longer than that of the FW. If we compare the results of FL and EL in FW and SW respectively, we can also draw the same conclusion, that is, TS-CNN (Single) is more suitable for the detection of long texts while TS-CNN (Multi) for short texts. We think this subtle difference between two proposed models comes from the different design patterns. TS-CNN(Single) pays more attention to the semantic coherence between sentences, while TS-CNN(Multi) to the expression of the internal semantics of each sentence.

\begin{figure*}[ht!]
  \centering
  \includegraphics[width=0.24\linewidth]{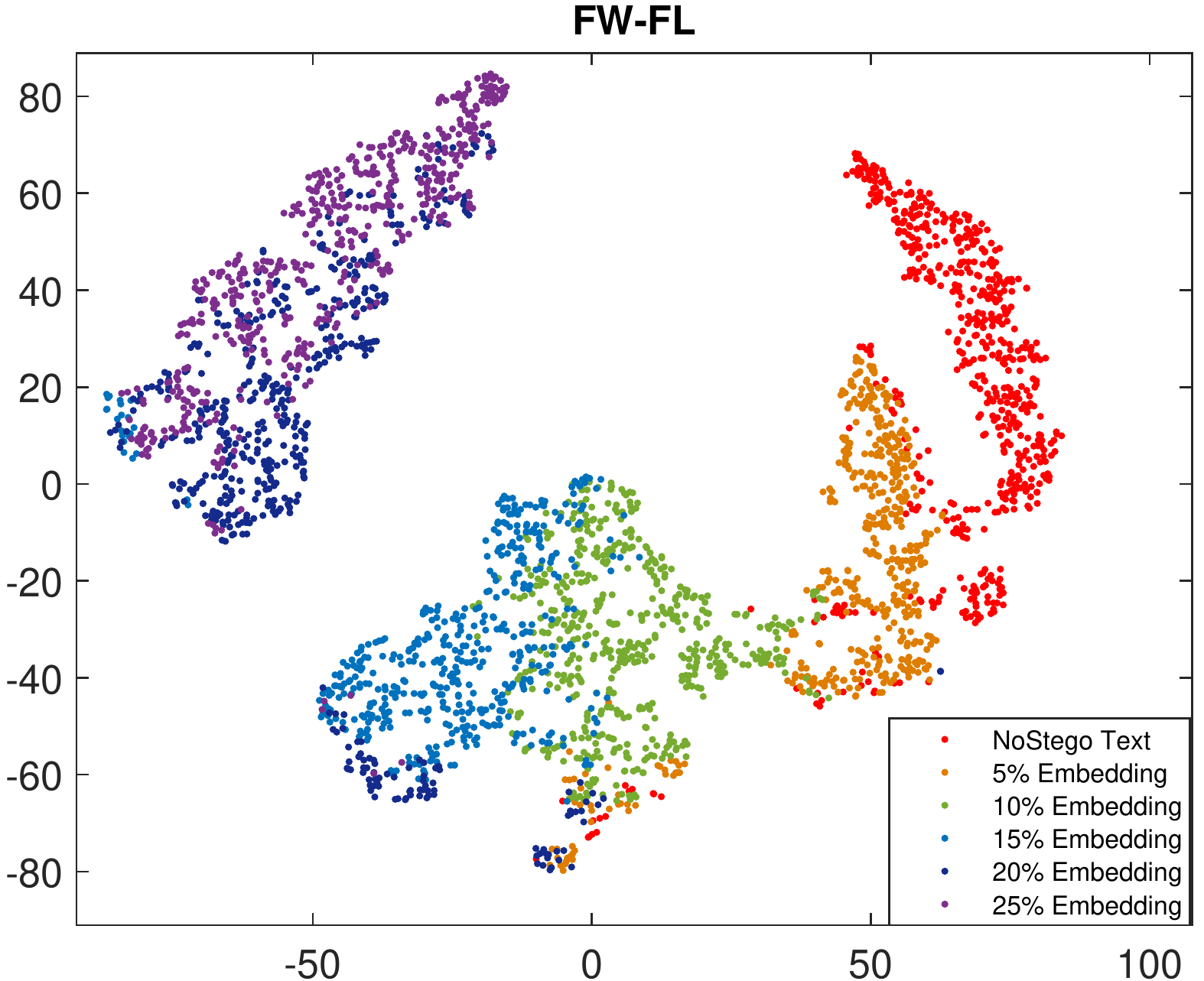}\hfill
  \includegraphics[width=0.24\linewidth]{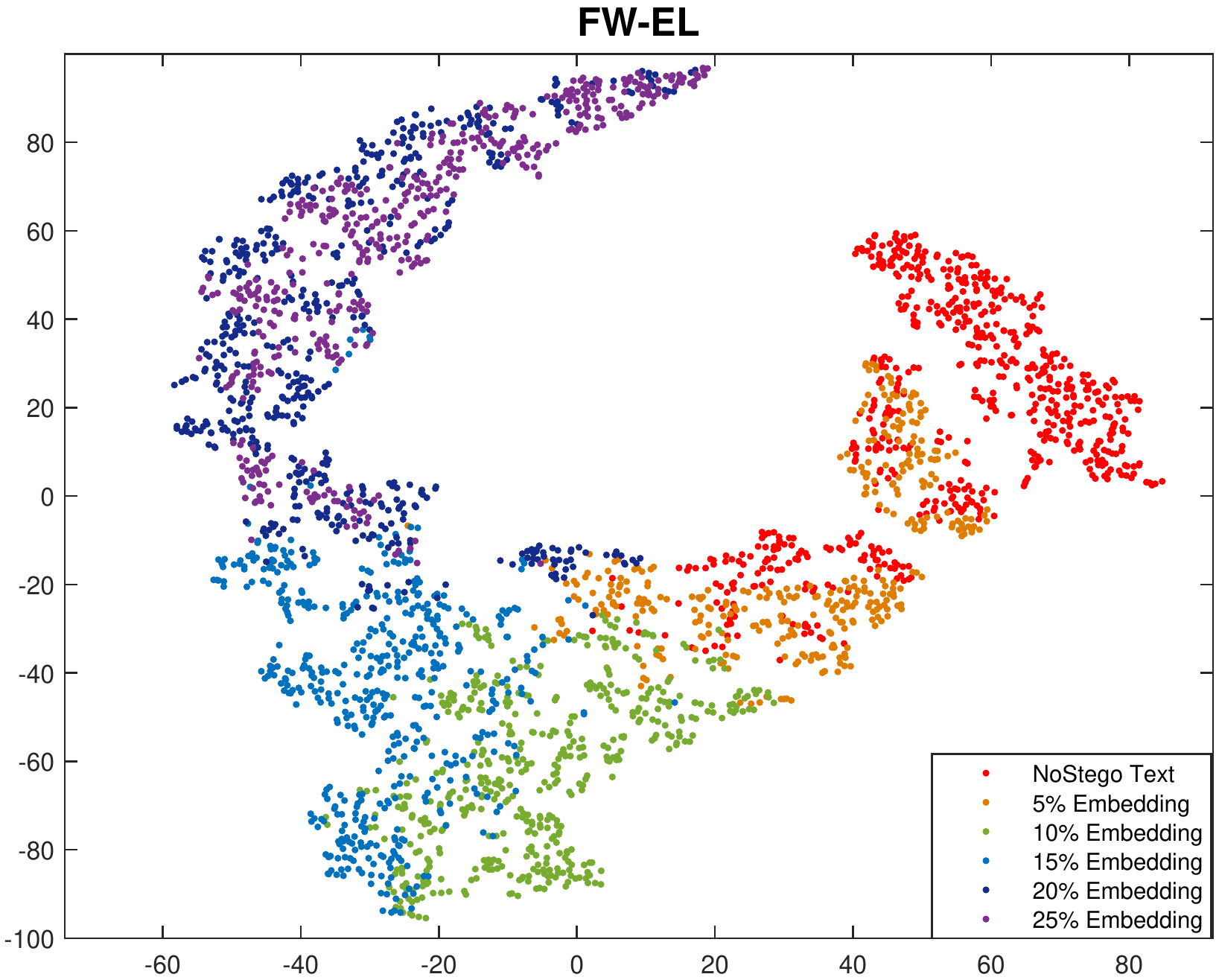}
  \includegraphics[width=0.24\linewidth]{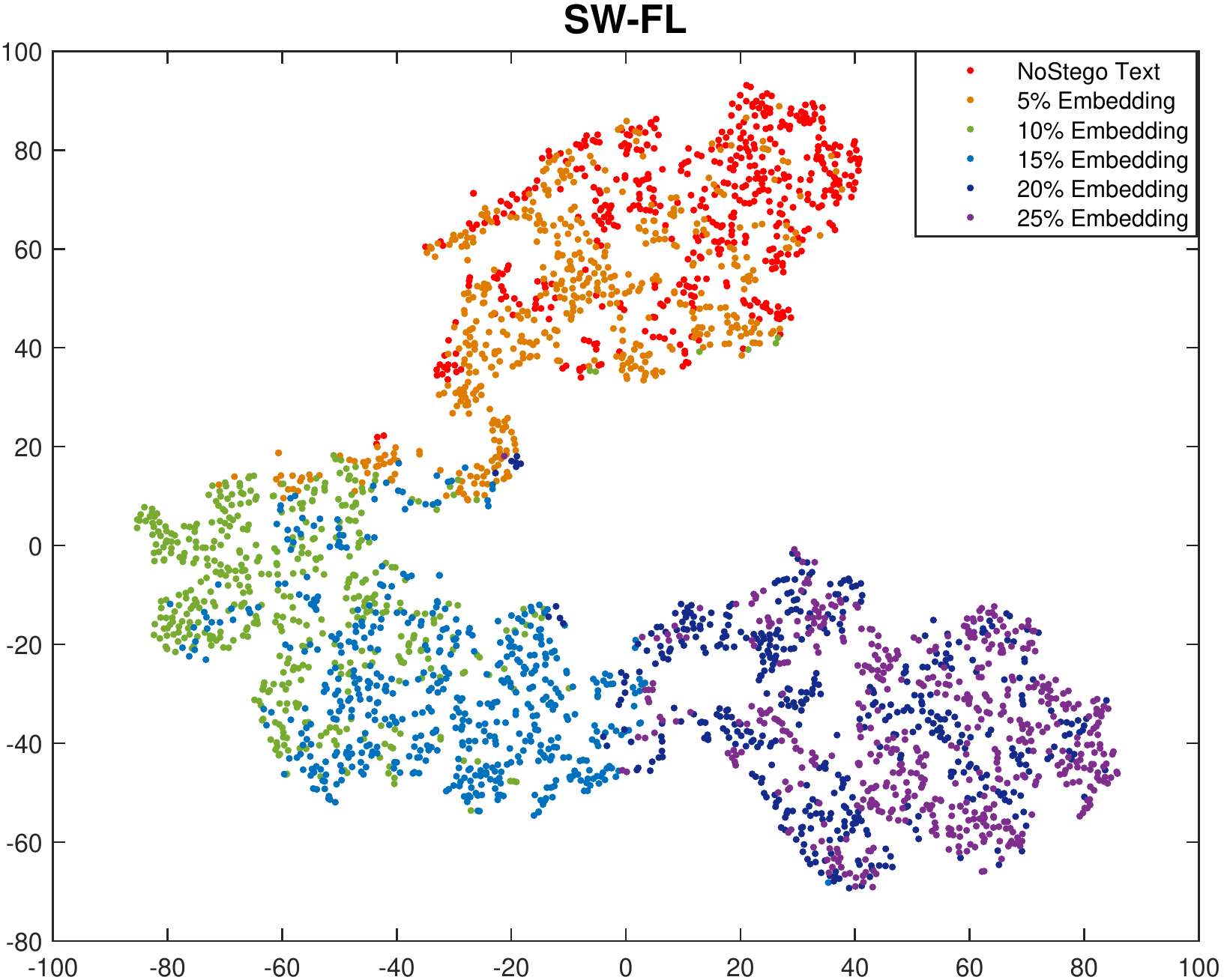}\hfill
  \includegraphics[width=0.24\linewidth]{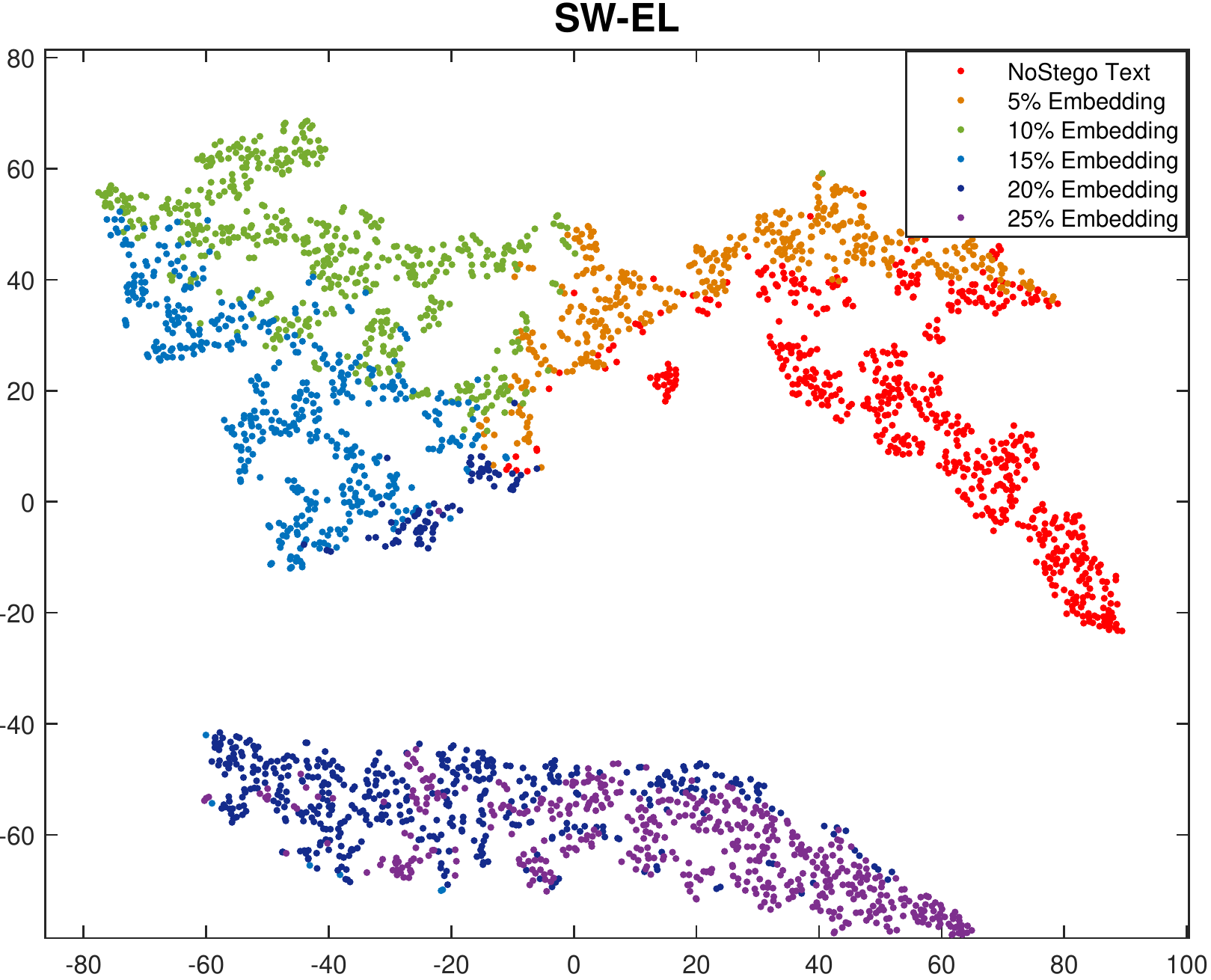}
  \caption{The distribution of texts under different information embedding rates in the semantic space. Each point represents a poem containing hidden information with different embedding rates from 0\% to 25\%, and different colors indicate different embedding rate poems. We can clearly see that as the embedded rate of hidden information increases in the generated texts, their distribution in the semantic space will gradually change and migrate.}
\label{fig:9}
\end{figure*}

\begin{table}[h]
\renewcommand\arraystretch{1.5}
\centering
\begin{tabular}{cc|c|c|c|c}
\toprule[1.5pt]
\multicolumn{2}{c|}{Model} &\multicolumn{2}{|c|}{FW} &\multicolumn{2}{|c}{SW}\\
\hline
\multicolumn{2}{c|}{Dataset} &FL &EL &FL &EL\\
\hline
\multirow{3}{*}{\cite{Meng2009Linguistic}} &{P} &0.258 &0.266 &0.261 &0.246\\
\cline{2-6}
&{R} &0.297 &0.311 &0.292 &0.286\\
\cline{2-6}
&{$F_1$} &0.245 &0.258 &0.253 &0.239\\
\hline
\multirow{3}{*}{\cite{samanta2016real}} &{P} &0.465 &0.510 &0.493 &0.540\\
\cline{2-6}
&{R} &0.473 &0.511 &0.492 &0.521\\
\cline{2-6}
&{$F_1$} &0.453 &0.503 &0.485 &0.533\\
\hline
\multirow{3}{*}{\cite{din2015performance}} &{P} &0.465 &0.513 &0.496 &0.568\\
\cline{2-6}
&{R} &0.472 &0.515 &0.501 &0.568\\
\cline{2-6}
&{$F_1$} &0.467 &0.514 &0.498 &0.566\\
\hline
\hline
\multirow{3}{*}{TS-CNN(Single)} &{P} &0.711 &0.744 &0.724 &0.772\\
\cline{2-6}
&{R} &0.708 &0.743 &0.735 &0.769\\
\cline{2-6}
&{$F_1$} &0.709 &0.743 &0.732 &0.769\\
\hline
\multirow{3}{*}{TS-CNN(Multi)} &{P} &0.751 &0.724 &0.737 &0.718\\
\cline{2-6}
&{R} &0.749 &0.718 &0.734 &0.712\\
\cline{2-6}
&{$F_1$} &0.748 &0.716 &0.732 &0.710\\
\bottomrule[1.5pt] 
\end{tabular}
\caption{\label{tab:4}The results of the proposed models' estimate of the capacity of the covert information in texts.}

\end{table}

Finally, if we only compare the performance of the two models we proposed, we find that TS-CNN (Multi) has better performance for short texts at low embedding rates, while TS-CNN (Single) shows better performance when the embedding rate is high, like for EL in SW, when the embedding rate is greater than 15\%, it can even achieve 100\% accuracy and recall. The reason is that the semantic expression of poetry consists of two parts, one is the semantic expression of each sentence itself, and the other is the semantic relevance between sentences. The samples in CT-Steg dataset are generated and written using the model proposed in \cite{Luo2017Text}, which uses an attention-based bidirectional encoder-decoder model to reinforce the correlation between sentences. Therefore, when embedding covert information in the generated poems and the embedding rate is low, it first damages the semantic expression of single sentences in poems, but still maintains a strong correlation between sentences. As the embedding rate increases, it gradually affects the semantic expression between sentences. While TS-CNN (Multi) extracts features from each sentence in the poems, TS-CNN (Single) pays more attention to the overall semantic expression, that is, the relevance between sentences.

\subsubsection{Embedded Rate Estimation}

As we have mentioned before, our model automatically extract high-level semantic features from texts and map them to a high-dimensional feature space (usually hundreds to thousands of dimensions). We can use t-Distributed Stochastic Neighbor Embedding (t-SNE)\cite{Maaten2014Accelerating} technique for the dimensionality reduction and visualization of these high-dimensional feature vectors, which can be found in Figure \ref{fig:9}. In this feature space, each point represents a poem containing hidden information with different embedding rates from 0\% to 25\%, and different colors indicate different embedding rate poems. 

From the Figure \ref{fig:9}, we find that points with the same embedding rate are clustered in the same area, indicating that our model accurately detects the subtle differences in the semantics of these texts under different information embedding rates. Moreover, as the embedded rate of hidden information increases, their distribution in the semantic space will gradually change and migrate, from the red area (which contains no hidden information) to the purple area(with 25\% covert information). When the embedding rate is low, such as 5\%, the area formed by the corresponding yellow dots and the red dots (containing no hidden information) are still have some overlap, but the change of the center of gravity can be clearly seen. When the embedding rate is high, for example, more than 20\%, that is, the area formed by the points of dark blue and purple dots have very clear boundaries with the red dots area. 

The results in Figure \ref{fig:9} fully demonstrate our model's ability in extracting semantic information with covert information, as well as the ability to implement text steganalysis using subtle differences in the semantic space distribution of text under different information embedding rates. This is also the core point of this paper. After embedding hidden information in texts, it will definitely damage the semantic expression of the texts. By using methods (such as TS-CNN) to extract the high-level semantic features of texts, we can use these subtle differences to achieve effective steganalysis.

Besides steganographically determined, we find our model can even make use of the distribution of input text in semantic space to estimate the capacity of hidden information inside. We mixed texts with covert information of different embedded rates, and made a set of experiments to test whether our model can estimate accurately. The experimental results are shown in Table \ref{tab:4}.

From Table \ref{tab:4}, we can see that our model can achieve an estimated accuracy of more than 70\% for the hidden information in the text, outperforms all the other models. Although there is still much room for improvement, we must note that there is no work can estimate the amount of covert information in texts. Therefore, our results will be much meaningful. We expect that there will be more relevant work in this part, and the performance will be further improved.

\section{Conclusion}

In this paper, we propose a text steganalysis method based on semantic analysis called TS-CNN. It uses convolutional neural network(CNN) to extract high-level semantic features of texts, and finds the subtle differences in the semantic space before and after embedding the secret information to achieve high efficiency and high accuracy detection of steganographic texts. For a piece of text input, TS-CNN will first extract its high-level semantic feature, and map it into high-dimensional semantic space. By analyzing its distribution in the semantic space, it will determine whether it contains covert information. To train and test the proposed model, we collected and released a large text steganalysis (CT-Steg) dataset, which contains a total number of 216,000 texts with various lengths and various embedding rates. Experimental results show that the proposed model can achieve nearly 100\% precision and recall, outperforms all the previous methods. Besides, our model can even make use of the distribution of input text in semantic space to estimate the capacity of the hidden information inside and the estimated accuracy rate is above 70\%. These results strongly support that using the subtle changes in the semantic space before and after embedding the secret information to conduct text steganalysis is feasible and effective. We hope that this paper will serve as a reference guide for researchers to facilitate the design and implementation of better text steganalysis.



\end{document}